\documentclass[twocolumn,twocolappendix,trackchanges]{aastex63}
\usepackage{lineno}
\usepackage{threeparttable}
\usepackage{appendix}
\usepackage{natbib}
\usepackage{color}
\usepackage{mathptmx}
\usepackage{amsmath}
\usepackage{url}
\usepackage{multirow}
\usepackage{verbatim}
\usepackage[utf8]{inputenc}

\newcommand{\msun}{\,M_\odot}

\graphicspath{{./}}

\begin{document}
	
	\shorttitle{DMDGs and Their Member Star Clusters Form Simultaneously in Galaxy Collisions}
	\shortauthors{Joohyun Lee, Eun-jin Shin, and Ji-hoon Kim}
	
	\title{Dark Matter Deficient Galaxies And Their Member Star Clusters Form Simultaneously During High-velocity Galaxy Collisions In 1.25 pc Resolution Simulations}
	
	\correspondingauthor{Ji-hoon Kim}
	\email{me@jihoonkim.org}
	
	\author[0000-0001-8593-8222]{Joohyun Lee}
	\affiliation{Center for Theoretical Physics, Department of Physics and Astronomy, Seoul National University, Seoul 08826, Korea}
	
	\author[0000-0002-4639-5285]{Eun-jin Shin}
	\affiliation{Center for Theoretical Physics, Department of Physics and Astronomy, Seoul National University, Seoul 08826, Korea}
	
	\author[0000-0003-4464-1160]{Ji-hoon Kim}
	\affiliation{Center for Theoretical Physics, Department of Physics and Astronomy, Seoul National University, Seoul 08826, Korea}
	\affiliation{Seoul National University Astronomy Research Center, Seoul 08826, Korea}
	
	\begin{abstract}
		
		How diffuse dwarf galaxies that are deficient in dark matter—such as NGC1052-DF2 and NGC1052-DF4—formed is a mystery.
		Along with their luminous member globular clusters (GCs), the so-called dark matter deficient galaxies (DMDGs) have challenged observers and theorists alike.
		Here we report a suite of galaxy collision simulations using the adaptive mesh refinement code {\sc Enzo} with 1.25 pc resolution, which demonstrates that high-velocity galaxy collisions induce the formation of DMDGs and their star clusters (SCs) simultaneously.  
		With a numerical resolution that is significantly better than our previous study, we resolve the dynamical structure of the produced DMDGs and the detailed formation history of their SCs, which are possible progenitors of the DMDG's member GCs. 
		In particular, we show that a galaxy collision with a high relative velocity of $\sim 300\;{\rm km\;s}^{-1}$, invoking severe shock compression, spawns multiple massive SCs ($M_\star \gtrsim10^6\msun$) in $<150\;{\rm Myr}$ after the collision.
		At the end of the $\sim 800 \;{\rm Myr}$ evolution in our fiducial run, the resulting DMDG of $M_\star \simeq 3.5\times10^{8} \msun$ hosts 10 luminous ($M_{V}\lesssim-8.5$ mag), gravitationally bound SCs with a line-of-sight velocity dispersion $11.2\;{\rm km\;s}^{-1}$.  
		Our study suggests that DMDGs and their luminous member SCs could form simultaneously in high-velocity galaxy collisions  while being in line with the key observed properties of NGC1052-DF2 and NGC1052-DF4.  
		
	\end{abstract}
	\keywords{galaxies: formation --- galaxies: evolution --- galaxies: star formation --- galaxies: star clusters: general --- cosmology: theory --- methods: numerical }

	\vspace{-2mm}
	
	\section{Introduction} \label{sec:intro}
	
	At the center of the intense debate on the formation mechanism that can address abnormal lack of dark matter and unusually bright populations of globular clusters (GCs) are two recently discovered ultra-diffuse dwarf galaxies, NGC1052-DF2 and NGC1052-DF4 (hereafter DF2 and DF4, respectively). 
	The spectroscopic study of their member GCs has revealed surprisingly small velocity dispersions compared to what is normally expected by their observed stellar masses.  
	This implies that DF2 and DF4 have very small dark-matter-to-stellar-mass ratios \citep[dark matter deficient galaxy (DMDG);][]{vanDokkum2018, vanDokkum2019}.\footnote{
    	This anomalous characteristic is based on the distance measurement, the accuracy of which is still under debate (see e.g., \citealt{Trujillo2019a} and \citealt{Trujillo2019b} for details, but also \citealt{vanDokkum2018c}, \citealt{Danieli2020} and \citealt{Shen2021a}).}
	The exceptionally large population of their luminous member GCs is also a mystery.  
	The GC luminosity function (GCLF) of DF2 peaks at $M_V\sim-9$, approximately 1.5 mag shifted to the brighter side from the Milky Way's GCLF \citep{vanDokkum2018b, Ma2020, Shen2021b}, indicating a large number of massive GCs. 
	Extensive spectroscopic studies of DF2 have found that their stellar population is $\sim$9 Gyr old on average and that the metallicity of the main stellar body differs from that of its member GCs \citep{Fensch2019, RuizLara2019}.
	As for the kinematics, recent studies hint that DF2 is likely a prolate rotator \citep{Emsellem2018, Lewis2020}.
	
	\begin{deluxetable*}{cccccccc}[ht!]
		\tablenum{1}
		\vspace{-1mm}
		\tabletypesize{\footnotesize}
		\tablecaption{\footnotesize A Suite of Idealized Galaxy Gollision simulations listed with their initial configurations and the stellar mass of the most massive DMDG formed}
		\vspace{-1mm}
		\tablewidth{0pt}
		\tablehead{
			\colhead{Run Name} & \colhead{Stellar Feedback Energy} & \colhead{Initial Gas Metallicity} & \colhead{Disk Inclination Angle} & \colhead{Disk Spin Relative Angle} & \colhead{$M_{\star,\,\rm DMDG}$} & \colhead{$t_{\rm end}$}\\[-2mm]
			\colhead{} & \colhead{(10$^{51}$ erg/150$\msun$ of Stars)} & \colhead{($Z_{\odot}$)}  & \colhead{($\theta_1$,$\phi_1$),$\,\,\,$($\theta_2$,$\phi_2$)} &  & \colhead{($10^8\msun$)} & \colhead{(Myr)}\\[-2mm]
			\colhead{(1)} & \colhead{(2)}  & \colhead{(3)} & \colhead{(4)} & \colhead{(5)} & \colhead{(6)} & \colhead{(7)}}
		\startdata
		{\tt Fiducial}\tablenotemark{\scriptsize \textcolor{red}{\textdagger}} & 1 & 1 & (0$^{\circ}$, 0$^{\circ}$),$\,\,\,$(0$^{\circ}$, 0$^{\circ}$) & 0$^{\circ}$ & 3.47 (3.72) & 800 (250)\\
		{\tt FB3} & 3.3 & 1 & (0$^{\circ}$, 0$^{\circ}$),$\,\,\,$(0$^{\circ}$, 0$^{\circ}$) & 0$^{\circ}$ & 2.76 & 250\\
		{\tt FB7} & 6.7 & 1 & (0$^{\circ}$, 0$^{\circ}$),$\,\,\,$(0$^{\circ}$, 0$^{\circ}$) & 0$^{\circ}$ & 1.46 & 250\\
		{\tt FB10} & 10 & 1 & (0$^{\circ}$, 0$^{\circ}$),$\,\,\,$(0$^{\circ}$, 0$^{\circ}$) & 0$^{\circ}$ & - & 50\\
		{\tt Low-metal}\tablenotemark{\scriptsize \textcolor{red}{\textdaggerdbl}} & 1 & $10^{-2}$ & (0$^{\circ}$, 0$^{\circ}$),$\,\,\,$(0$^{\circ}$, 0$^{\circ}$) & 0$^{\circ}$ & 0.90 & 250\\
		{\tt Retrograde} & 1 & 1 & (0$^{\circ}$, 0$^{\circ}$),$\,\,\,$(180$^{\circ}$, 0$^{\circ}$) & 180$^{\circ}$ & 1.76 & 250\\
		{\tt Tilt-1} & 1 & 1 & ($-22\rlap{.}^{\circ}5$, 0$^{\circ}$),$\,\,\,$($22\rlap{.}^{\circ}5$, 0$^{\circ}$) & 45$^{\circ}$ & 2.33 & 250\\
		{\tt Tilt-1-FB10} & 10 & 1 & ($-22\rlap{.}^{\circ}5$, 0$^{\circ}$),$\,\,\,$($22\rlap{.}^{\circ}5$, 0$^{\circ}$) & 45$^{\circ}$  & - & 50\\
		{\tt Tilt-2} & 1 & 1 & (0$^{\circ}$, $-22\rlap{.}^{\circ}5$),$\,\,\,$(0$^{\circ}$, $22\rlap{.}^{\circ}5$) & 45$^{\circ}$  & - & 50 \\
		{\tt Tilt-2-FB10} & 10 & 1 & (0$^{\circ}$, $-22\rlap{.}^{\circ}5$),$\,\,\,$(0$^{\circ}$, $22\rlap{.}^{\circ}5$) & 45$^{\circ}$  & - & 50\\
		\enddata
		\tablecomments{\scriptsize
			The simulation configurations and the stellar mass of the most massive DMDG formed:
			(1) run name,
			(2) stellar feedback energy, 
			(3) initial metallicity of the gas disks, 
			(4) angle by which each of the two colliding galactic disks is rotated about the +$x$-axis ($\theta_1$, $\theta_2$) and about the +$y$-axis ($\phi_1$, $\phi_2$), starting from its original position in the $x$-$y$ plane (see the top-left panel of Figure \ref{fig:1}),
			(5) relative angle between the angular momentum vectors of the two galactic disks,
			(6) stellar mass within 12 kpc from the center of the most massive DMDG (``-'' $=$ no DMDG formed), 
			(7) time since the pericentric approach of the two progenitor disks when we end the simulation.}
		\begin{tablenotes}
			\item\textdagger$\,${\scriptsize This is the fiducial run presented in all of the figures in this article. See Sections \ref{sec:method} and \ref{sec:3} for more information.}
			\item\textdaggerdbl$\,${\scriptsize This is the run presented in the right panels of Figure \ref{fig:5} in Section \ref{sec:3.2}.}
			\tabletypesize{\footnotesize}
		\end{tablenotes}
		\vspace{-8mm}
		\label{tab:1}
	\end{deluxetable*}
	
	These observations could provide a testbed for various formation scenarios proposed for DF2 and DF4.
	In one of such scenario suggested by \cite{Silk2019}, a ``mini-Bullet cluster-like'' event on a galaxy scale decouples dark matter from dissipative baryons.\footnote{
    	Other scenarios include the tidal dwarf galaxies \citep{{Fensch2019}} and the dwarf galaxies with tidally stripped dark matter halos \citep{Ogiya2018, Jackson2021}. See \cite{Shin2020} for more information.}
	The high-velocity galaxy collision induces a strong shear and shock compression in the interstellar medium (ISM), instigating the formation of DMDGs and their bright member GCs \citep{Silk2019, Trujillo2020}.
	Numerical studies have demonstrated the formation of young massive clusters (YMCs) and/or GC candidates triggered by galaxy mergers in idealized simulations \citep{Li2004, Bournaud2008, Kruijssen2011, Kruijssen2012, Hopkins2013, Renaud2015, Maji2017, Moreno2019, Lahen2019, Lahen2020} and in cosmological simulations \citep{Li2017, Kim2018, Ma2020b}.
	To validate the scenario by \cite{Silk2019}, we previously simulated idealized galaxy collisions using two different gravito-hydrodynamics codes and showed that $\sim$10$^{8}\msun$ DMDGs are produced when two gas-rich, dwarf galaxies collide with a relative velocity of $\sim 300\;{\rm km\;s}^{-1}$ \citep{Shin2020}.  
	However, the chosen resolution of 80 pc limited our ability to examine the inner structure of the resulting DMDGs, or to resolve the formation of their member SCs.  
	
	In this Letter, we examine the galaxy collision–induced DMDG formation scenario using adaptive mesh refinement (AMR) simulations with significantly improved 1.25 pc resolution.    
	We not only demonstrate the DMDGs' formation but also explore the detailed formation histories and the dynamical properties of the DMDG and their member SCs which could eventually evolve into bright GCs.

	\section{Simulations} \label{sec:method}
	
	The initial condition used for this study is similar to what was adopted in \citet{Shin2020} but with greatly improved particle resolution: a dark matter halo of mass $3.88\times 10^{9}\msun$ with a particle mass $1.55\times10^{3}\msun$ each, a stellar disk of $4.21\times 10^{8}\msun$ with a particle mass $1.05\times 10^3\msun$ each, and a gas disk of $1.68\times 10^{9}\msun$.
	The initial gas fraction is thus $f_{\rm gas}=M_{\rm gas}/M_{\rm galaxy}=0.28$, but because $\sim30\%$ of the gas is consumed before the collision, $f_{\rm gas}$ becomes $\sim 0.2$ when two galaxies collide at $t=0$. 
	Our {\tt Fiducial} run is a co-planar prograde–prograde collision of the two identical galaxies with a relative velocity of $\sim 300\;{\rm km\;s}^{-1}$ at a 40 kpc distance, and with a pericentric distance of 1 kpc (see Table \ref{tab:1}).\footnote{
    	Readers should note that this initial condition models a rare type of hyperbolic galaxy collision events, but a type that does exist in a simulated universe such as IllustrisTNG \citep[for details, see][]{Shin2020}.}
	By varying the stellar feedback energy, initial gas metallicity, disk inclinations and spins, we form a suite of numerical experiments.  
	The simulation configurations and the stellar masses of the resulting DMDGs are listed in Table \ref{tab:1}.
	
	\begin{figure*}[t]
		\centering
		\vspace{-1mm}  
		\includegraphics[width=0.86\textwidth]{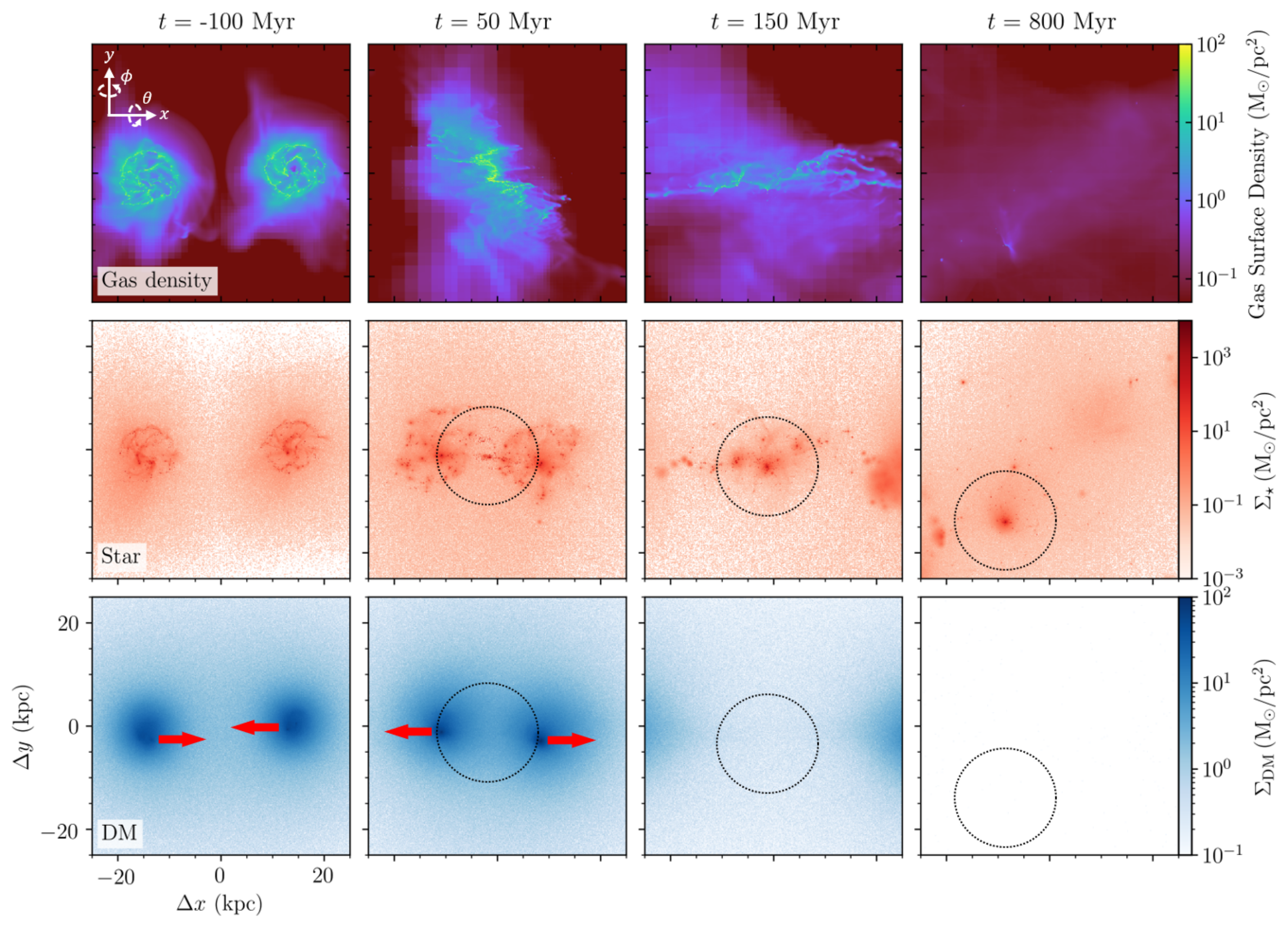}
		\vspace{-3mm}
		\caption{ 
			$t=-100$, 50, 150, 800 Myr snapshots of the fiducial, idealized collision of two identical gas-rich dwarf galaxies of $M_{200}=5.95\times10^9\msun$ each, and the resulting DMDGs (the {\tt Fiducial} run in Table \ref{tab:1}).
			Surface densities of gas (top row), stars (middle row) and dark matter (bottom row). 
			$t=0$ is set to the moment when the two progenitor galaxies are at pericentric approach.    
			The co-planar galaxy collision (i.e., both progenitor disks lying in the $x$-$y$ plane) with a relative velocity of 300 ${\rm km\;s}^{-1}$ produces multiple DMDGs with stellar mass $\sim$10$^{8}\msun$ each. 
			The dotted circles of 10 kpc radius are centered on the most massive DMDG produced.    
			See Section \ref{sec:3.1} for more information. 
		}
		\label{fig:1}
		\vspace{2mm}
	\end{figure*}
	
	The simulations are performed with the publicly available code {\sc Enzo} \citep{Bryan2014, Brummel-Smith2019} with the {\sc Zeus} hydrodynamics solver  \citep{Stone1992a, Stone1992b}.
	Radiative gas cooling and heating rates along with the metagalactic ultraviolet (UV) background are computed by the {\sc Grackle} library that interpolates the lookup table from {\sc Cloudy} \citep{Smith2017}.
	To describe feedback-regulated star formation, we employ the star-forming molecular cloud model \citep[SFMC; for details, see][]{Kim2013, Kim2019}, a change from a simpler prescription in \citet{Shin2020}.
	A particle representing a SFMC forms when {\it (1)} the density of a gas cell exceeds $4\times 10^4$ cm$^{-3}$, {\it (2)} the gas flow is converging, {\it (3)} the cooling time of the cell is shorter than its dynamical time ($t_{\rm dyn}$), and {\it (4)} the created SFMC particle is heavier than $M_{\rm SFMC} = 10^3 \msun$.
	Depicting inefficient star formation in molecular clouds \citep{Krumholz2007}, the SFMC particle returns 80\% of its original mass in 12 $t_{\rm dyn}$,\footnote{
    	The local star formation efficiency (SFE) per freefall time ($\epsilon_{\rm ff} \sim 1\%$) is based on \cite{Kim2011, Kim2019}. Other studies have also found that varying the SFE parameter by not more than an order of magnitude does not significantly affect the SC properties \citep{Li2020, Ma2020b}.}
     together with thermal energy peaking at 1 $t_{\rm dyn}$ (four different energy choices for {\tt Fiducial} to {\tt FB10} runs between 10$^{51-52}$ erg per 150$\msun$ of stars formed) and metals \citep[2\% of the returned gas; see also][]{Cen1992, Kim2011}.\footnote{
    	 While it is possible that the momentum imparted by supernovae could be underestimated in our thermal feedback model, the subgrid physics implementation adopted here have consistently shown that the the star formation is efficiently self-regulated and the DMDGs are formed, across multiple resolution (resolution study at 80, 5, 1.25, and 0.625 pc) and feedback strengths ({\tt Fiducial}, {\tt FB3}, and {\tt FB7} run in Table \ref{tab:1}).} \label{footnote:feedback}
	
	\begin{figure*}[t]
		\centering
		\includegraphics[width=\textwidth]{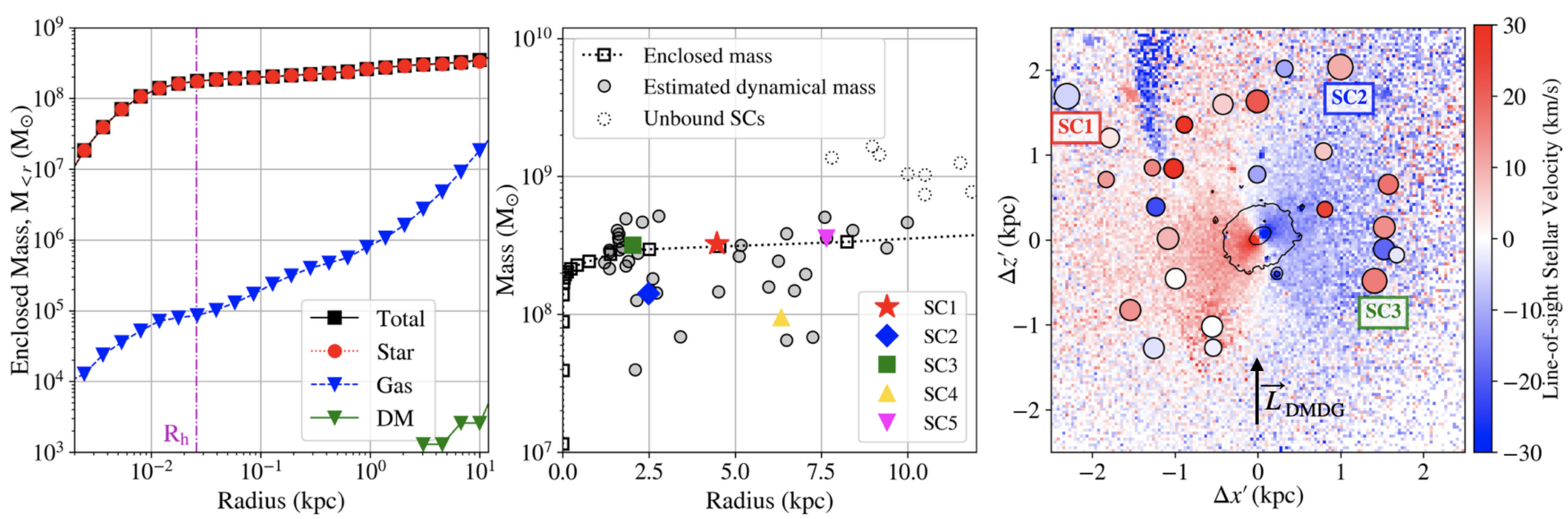}
		\vspace{-5mm}    
		\caption{
			Left panel: the radial profile of enclosed masses in a logarithmic scale from the center of the most massive DMDG at $t=800$ Myr in the {\tt Fiducial} run.
			The DMDG consists of mostly stars, lacking in dark matter and gas. 
			The vertical magenta line denotes the half-mass radius $R_{\rm h}$ of the DMDG.
			Middle panel: the dynamical mass within $r$ estimated from the individual SC's circular velocity, $M_{<}(r) =r\,v_{\rm circ}^2(r)/{\rm G}$, now in a linear scale plot (circles).  
			Among 42 gravitationally bound SCs the DMDG hosts, the five most massive SCs  are shown with colored markers.
			For comparison, the true curve of the enclosed total mass is also included (dotted line with squares).    
			Right panel: the line-of-sight stellar velocities in a (5 kpc)$^2$ box centered on the DMDG at $t\,=\,800$ Myr, projected to a plane whose vertical coordinate ($z^{\,\prime}$) is aligned with the total angular momentum of the DMDG (\textbf{\textit{L}}$_{\rm DMDG}$).
			Colored circles denote the location of SCs and their line-of-sight velocities with radii of $2\,(M_{\rm SC}/\msun)^{1/3}$ kpc. 
			Two isosurface density contours are also shown: ${\rm log[ \Sigma_{\star}/(\msun \, pc^{-2}) ]}= 0$ and 1.   
			See Section \ref{sec:3.1} for more information.
		}
		\label{fig:2}
		\vspace{2mm}        
	\end{figure*}
	
	A pair of colliding galaxies is initialized in a $(1.311 {\,\rm Mpc})^{3}$ box of $64^{3}$ root grid.  
	We adaptively refine the galaxies up to level $l_{\rm max} = 14$ achieving 1.25 pc resolution.\footnote{
    	We create a relaxed galactic disk by gradually increasing the resolution for 250 Myr, thereby suppressing artificial starbursts: maximum spatial resolution $= 80$ pc for the first 100 Myr ($l_{\rm max} = 8$), 20 pc for  the next 100 Myr ($l_{\rm max} = 10$), and 5 pc for the last 50 Myr ($l_{\rm max} = 12$). The relaxation step ends at $t=-150\;{\rm Myr}$, the beginning of our 1.25 pc resolution simulation discussed in this article ($t=0$ is when the two galaxies are at pericentric approach; see, e.g., Figure \ref{fig:4}). During the relaxation, SFMC formation is either not allowed (i.e., dense gas clumps are supported only by the Jeans pressure floor) or allowed with a coarser-resolution recipe.}
	After the two progenitor galaxies pass each other ($t=100$ Myr), we refine the cells only in a pre-defined region that only includes the produced DMDGs. 
	This {\tt RefineRegion} is initially set to a (16 kpc)$^3$ box containing the most massive DMDG throughout the run, and is surrounded by static nested boxes of cascading resolution from level 0 to 8.
	At level $l$, a cell splits into eight child cells whenever the cell contains more gas  than $M^{l}_{\rm ref,\,gas}=2^{-0.554 (l-14)}\times M^{14}_{\rm ref,\,gas}$ or more particles than $M^{l}_{\rm ref,\,part}=2^{-0.387 (l-14)}\times M^{14}_{\rm ref,\,part}$ where $M^{14}_{\rm ref,\, gas}=2000\;{\rm M_\odot} \simeq 2\; M_{\rm Jeans}^{180 \,{\rm K}}$ and $M^{14}_{\rm ref,\,part}=4000\;{\rm M_\odot} = 4\;M_{\rm SFMC}$.

	\section{Results} \label{sec:3}
	
	\subsection{Collision-induced DMDGs: Kinematics and Morphology} \label{sec:3.1}
	
\begin{figure*}[t] 
		\centering
		\vspace{-1mm}        
		\includegraphics[width=0.86\textwidth]{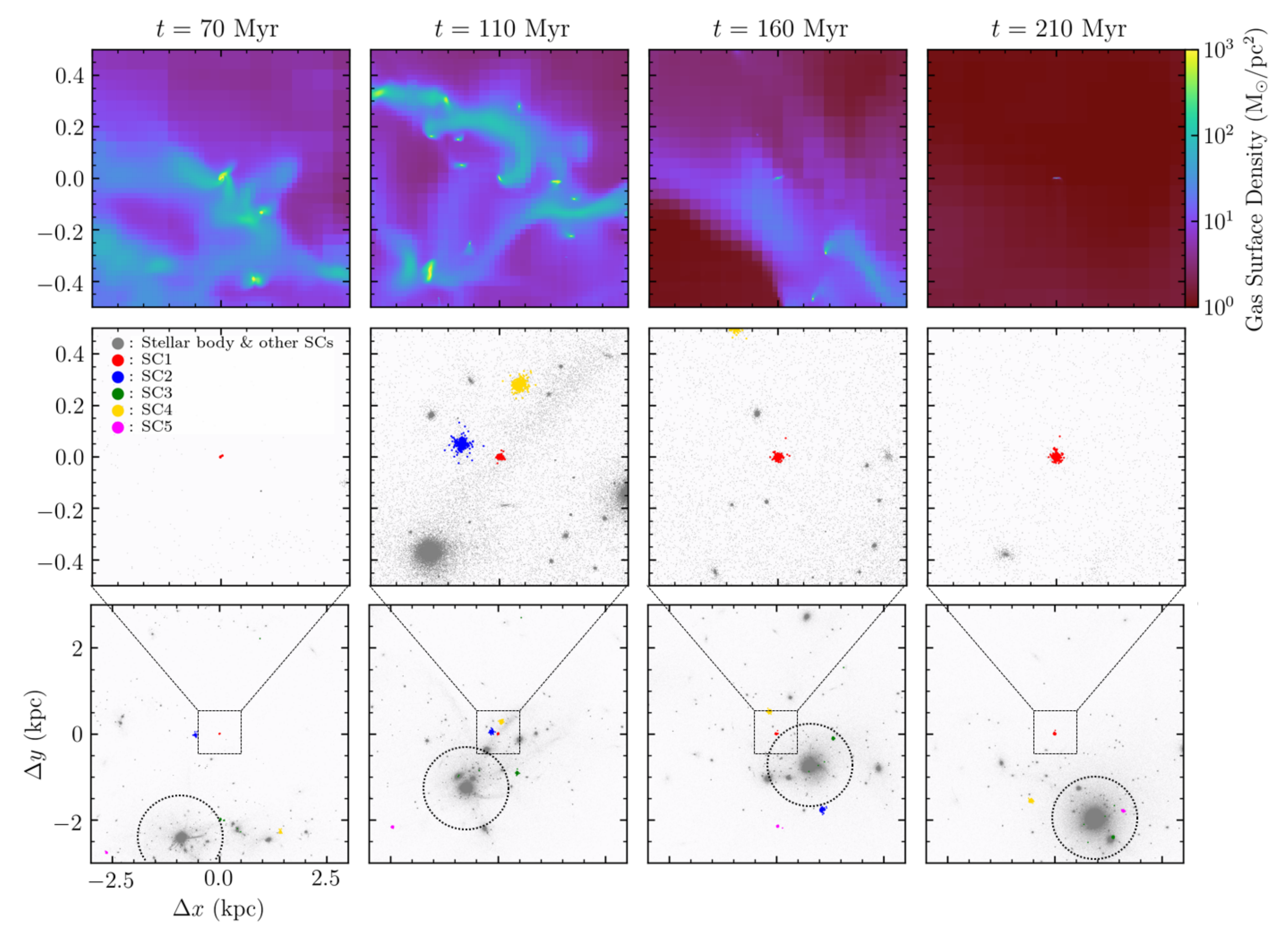}
		\vspace{-3mm}       
		\caption{
			Similar to Figure \ref{fig:1}, but zoomed in on the most massive SC (``SC1'') created around the most massive DMDG in the {\tt Fiducial} run. 
			$t=$ 70, 110, 160, and 210 Myr snapshots of the gas surface density and stellar distribution in a (1 kpc)$^2$ box (top and middle rows), and in a (6 kpc)$^2$ box (bottom row).
			The red dots in the middle and bottom panels denote the stars in SC1  (the same color codes and SCs  as in Figure \ref{fig:2}).
			The dotted circle of 1 kpc radius marks the location of the DMDG in each of the bottom panels.
			See Section \ref{sec:3.2} for more information.
		}
		\label{fig:3}
		\vspace{2mm}        
	\end{figure*}

	Figure \ref{fig:1} shows the collision sequence of two gas-rich dwarf galaxies approaching with a relative velocity of $\sim 300\;{\rm km\;s}^{-1}$ (the {\tt Fiducial} run in the first row of Table \ref{tab:1}; similar to what was extensively tested in \citealt{Shin2020}). 
	We define $t=0$ as the time when the two galaxies are at pericentric approach. 
	As shown in the top row, we find that extremely dense gas clumps form during $t=0-50\;{\rm Myr}$ as the colliding columns of gas undergo severe shock compression near the first contact point (the center of each panel) and along the tidally stripped gas streams.
	The highly compressed gas clumps of $\Sigma_{\rm gas} \sim 10^{3-4}\,{\rm M}_{\odot}\,{\rm pc}^{-2}$ (see also Figure \ref{fig:3})—a typical value for extremely dense giant molecular clouds—invokes the formation of DMDGs and their member SCs \citep{Kim2018, Lahen2020, Ma2020b}.
	Strong bursts of star formation continue inside the gas clumps during $t=50-150\;{\rm Myr}$ (middle row). 
	The newly formed stars remain near the first contact point and begin to form one or more gravitationally bound structures without the help of pre-existing dark matter potentials.
	In contrast, the two dark matter halos simply pass through each other during this high-velocity collision.  
	As a result, as seen in the bottom row of Figure \ref{fig:1}, hardly any dark matter particle remains in the resulting DMDGs. 
	Among the three DMDGs identified by the {\sc Hop} halo finder \citep{Efstathiou1985} in the {\tt Fiducial} run, the most massive one has $M_\star \simeq 3.5\times10^{8} \msun$ and almost no dark matter (see also Figure \ref{fig:2}). 
	In Figure \ref{fig:1}, the most massive DMDG found is marked with a dotted circle in each panel.  
	The DMDG drifts away slightly from the first contact point by $t=800\;{\rm Myr}$ (the fourth panels in each row). 
	
	In Figure \ref{fig:2}, we present the dynamical properties of the most massive DMDG produced in the {\tt Fiducial} run at $t=800\;{\rm Myr}$.
	The left panel displays the radial profiles of the enclosed masses in different components: stars, gas, and dark matter. 
	The DMDG primarily consists of stars and is nearly devoid of not only dark matter but also gas, which is in line with the reported gas-poor nature of DF2 \citep[$M_{\rm gas, \, DF2}\lesssim3.15\times10^6\msun$;][]{Chowdhury2019}.
	The half-mass radius $R_{\rm h}$ of the DMDG's stellar body is 0.026 kpc, which is $>10$ times smaller than that of a typical ultra-diffuse galaxy ($\sim 1.5 \;{\rm kpc}$) or that of DF2 \citep[$\sim 2.2 \;{\rm kpc}$;][]{vanDokkum2015, vanDokkum2018}.
	However, it is important to note that our simulated DMDGs evolved only for $< 800 \;{\rm Myr}$ in an idealized, isolated environment.  
	Some morphological properties of a compact stellar object, such as $M_{\rm gas}$ and $R_{\rm h}$ could be sensitive to the energetic supernovae feedback \citep{Trujillo2021}, or the tidal interaction with massive neighbors.\footnote{For example, the DMDGs formed in an isolated simulation may harbor more gas than DF2 does. However, \citet{Shin2020} demonstrated that if the simulated DMDG orbits around another massive galaxy, the tidal interaction could deprive the DMDG of most of its gas in a few Gyr.}
	$R_{\rm h}$ also changes frequently at this early stage of evolution as a number of new SCs in the outskirts leave the DMDG, or because of the dynamical relaxation or heating.
	
	We now study the kinematics of SCs  formed simultaneously with the DMDG that are identified by the {\sc Hop} algorithm reconfigured to locate SCs.
	We fit each SC's stellar density profile to a power law, $\rho (r)\,=\rho_{0}\,(1+r^{2}/a^{2})^{-{(\gamma +1)/2}}$ \citep{EFF1987}, and obtain its scale radius $a$. 
	$R=10\,a$ is adopted as the extent of each SC  when we determine its physical quantities such as mass, half-mass radius, orbital velocity, average metallicity, and creation time. 
	In the {\tt Fiducial} run, a total of 50 SCs are found within 12 kpc from the center of the DMDG at $t=800\;{\rm Myr}$; among them, 42 are gravitationally bound to the DMDG, and eight are unbound. 
	The middle panel of Figure \ref{fig:2} presents the dynamical mass within radius $r$ estimated from each SC's circular velocity, $M_{<}(r) =r\,v_{\rm circ}^2(r)/{\rm G}$.  
	Shown for comparison is the true curve of the DMDG's enclosed mass at  $r$.   
	It is notable that the dynamical masses estimated by the gravitationally bound SCs  lie close to the true enclosed mass.
	This implies that the gravitationally bound ``member'' SCs  of the DMDG are virialized, a key assumption when the dynamical masses of stellar systems are observationally evaluated.
	
	In the right panel of Figure \ref{fig:2}, we inspect the dynamical structure of the DMDG in the {\tt Fiducial} run.
	It displays the line-of-sight velocities of the stars and SCs in the DMDG, viewed from an angle that is perpendicular to the angular momentum vector of the DMDG (\textbf{\textit{L}}$_{\rm DMDG}$) at $t=800\;{\rm Myr}$.
	We first find that the line-of-sight velocity dispersion of the 42 gravitationally bound SCs  is $\sigma=\,12.6\;{\rm km\;s}^{-1}$. 
	The 10 most luminous bound SCs ($M_{V} < -8.5$; see Figure \ref{fig:6} and Section \ref{sec:3.2}) have $\sigma=\,11.2\;{\rm km\;s}^{-1}$.  
	The small spread in velocity—smaller than what is normally expected by its stellar mass—is an indication that the resulting galaxy has a small dark-matter-to-stellar-mass ratio.
	Our finding is in line with the observed value of $8.4\;{\rm km\;s}^{-1}$ for the luminous SCs  in DF2 \citep{vanDokkum2018} and $5.8\;{\rm km\;s}^{-1}$ in DF4 \citep{vanDokkum2019}, considering that the observation is conducted in an oblique plane.
	Second, from the ``butterfly-shaped'' line-of-sight velocities and the isosurface density contours, we find that the DMDG is an oblate rotator with a major-to-minor axis ratio $\sim 2$.
	The rotation velocity of the stars is $\sim15\;{\rm\,km\;s}^{-1}$ at $4\;R_{\rm h} < r < 10\;R_{\rm h}$ with  a noticeable velocity gradient along the major axis of the system at $r\lesssim R_{\rm h}$ ($v_{\rm circ, \, max}\sim 35\;\rm\,km\;s^{-1}$ at $r=25$ pc).
	For comparison, the stellar body of DF2 is reported to be a prolate rotator with a rotational velocity of $10.8 - 12.4\;{\rm km\;s}^{-1}$ (\citealt{Emsellem2018} and \citealt{Lewis2020}, respectively).
	
	However, by comparing the {\it specific} angular momentum of the DMDG (\textbf{\textit{l}}$_{\rm DMDG}$) and that of the system of 42 bound ``member'' SCs around the DMDG's center (\textbf{\textit{l}}$_{\rm SC}$), we discover that the SCs are not in a coherent rotation around the DMDG, nor preferentially in the disk plane of the DMDG. (i.e., $|$\textbf{\textit{l}}$_{\rm SC}|/|$\textbf{\textit{l}}$_{\rm DMDG}|=0.042$).
	The rotation axis of the system of SCs about the DMDG—if it exists at all—is not aligned with that of the whole DMDG (i.e., \textbf{\textit{l}}$_{\rm SC}$  and \textbf{\textit{l}}$_{\rm DMDG}$ makes a 69$^{\circ}$ angle). 
	The discrepancy in the dynamical states of the DMDG and the system of its member SCs could be attributed to their different formation processes (see Section \ref{sec:3.2}). 
	
	Lastly, using the suite of simulations that we performed (Table \ref{tab:1}), we remark on how the formation of the collision-induced DMDGs depends on the simulation parameters.
	{\it (1)} First, DMDGs form if a sufficiently large amount of gas participates in the supersonic collision and forms self-gravitating gas clumps.  
	For example, in the {\tt Tilt-1} run, although the collision is no longer co-planar, a portion of the disk gas along the $x$-axis (see Figure \ref{fig:1}) still clashes into a large column of gas in the other disk, producing two DMDGs.  
	In contrast, in the {\tt Tilt-2} run, the disk gas finds only a small column of gas in the other disk during its first near head-on collision, so no DMDG forms \citep[for a quantitative discussion, see Section 3.3 of][]{Shin2020}.
	{\it (2)} The strength of stellar feedback affects the internal and kinetic energy of the ISM and dictates the DMDG formation.
	In the {\tt FB10} run where the feedback is 10 times more energetic than in the {\tt Fiducial} run, clumps are easily dispersed via stellar feedback, with no DMDG formed.
	{\it (3)} The metallicity of the disk gas influences the DMDG formation.
	In the {\tt Low-metal} run where the initial gas metallicity is set to $10^{-2} \,Z_\odot$ (100 times lower than in the {\tt Fiducial} run), the relatively less metal-enriched gas slows down the gas cooling and suppresses star formation.  
	As a result, the mass of the resulting DMDG is only one-third of what is found in the {\tt Fiducial} run, hosting only nine SCs  (as opposed to 50 in the {\tt Fiducial} run; see also Figure \ref{fig:5}).
	{\it (4)} The velocity shear in the gas at the first contact surface affects the dynamical properties of the resulting DMDG.
	In the {\tt Fiducial} run of a prograde–prograde collision, the ISM experiences a strong shear as the colliding columns of gas move in the opposite directions.  
	Therefore, a DMDG is born in a clump that is in coherent rotation, resulting in a rotation-supported structure.    
	In contrast, in the {\tt Retrograde} run with a prograde–retrograde collision, the velocity vectors in the collapsing gas clumps are highly disordered, resulting in a DMDG with less rotation (i.e., $|$\textbf{\textit{l}}$_{\tt Retrograde}|/|$\textbf{\textit{l}}$_{\tt Fiducial}|\sim 0.1$).

	\subsection{SCs in the Collision-induced DMDGs: Formation History and Key Observables} \label{sec:3.2}
	
\begin{figure}[t] 
		\centering
		\vspace{-1mm}      
		\includegraphics[width=0.48\textwidth]{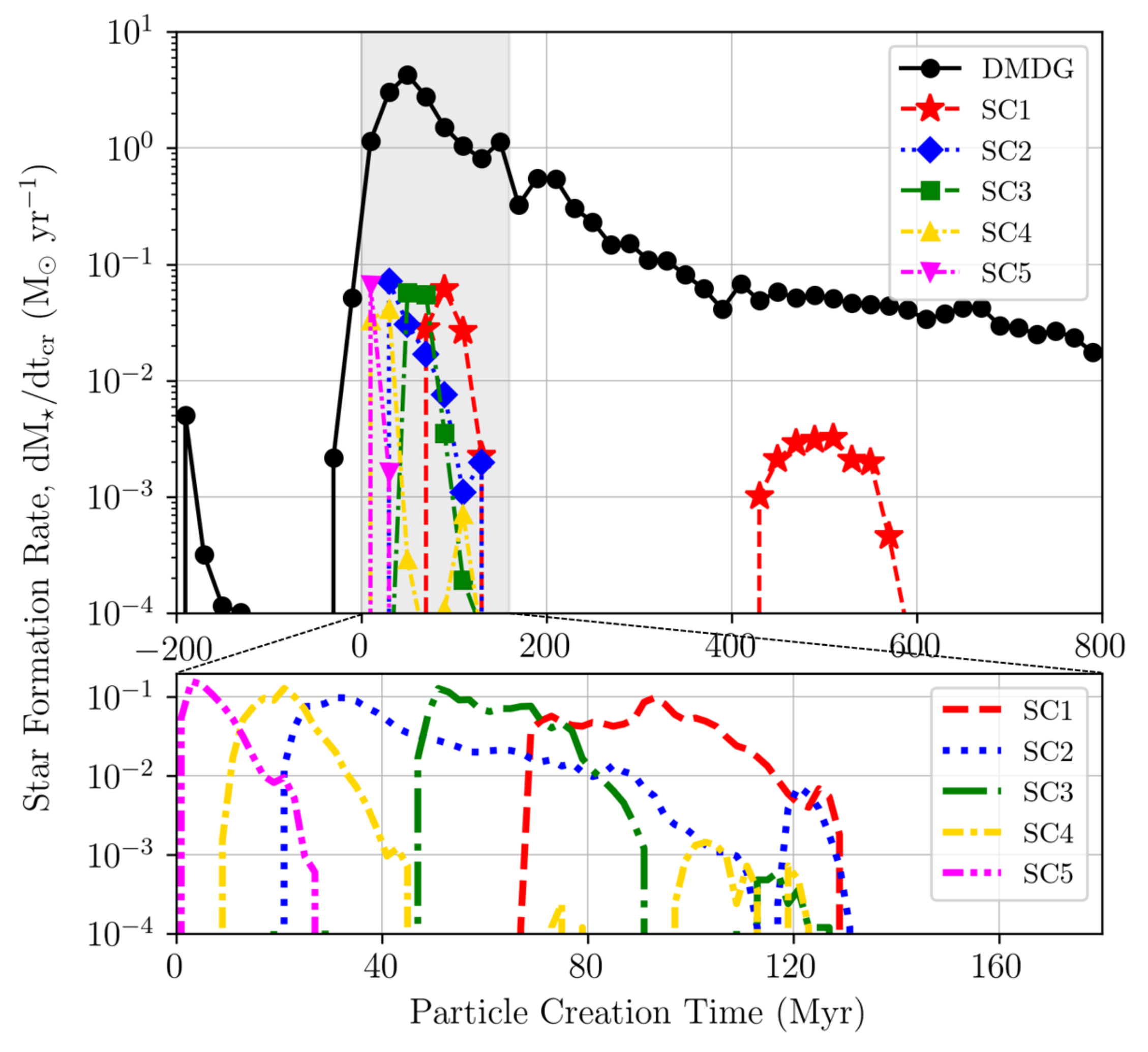}
		\vspace{-6mm}      
		\caption{
			Top panel: the star formation history in the most massive DMDG and its five most massive member SCs in the {\tt Fiducial} run (the same SCs and colored markers as in Figure \ref{fig:2}).
			Bottom panel: zoomed in on the time that the five SCs are created.
			The five SCs and most of the stars in the DMDG's stellar body form during $t=0-150$ Myr, immediately after the progenitor galaxies' pericentric approach.
			See Section \ref{sec:3.2} for more information.
		}  
		\label{fig:4}
		\vspace{0mm}      
	\end{figure}

	We turn our attention to the formation history and the properties of SCs in the collision-induced DMDGs.
	Figure \ref{fig:3} again shows the collision sequence of the {\tt Fiducial} run, but zooming in on the site where the most massive SC of the most massive DMDG was formed (``SC1''; red dots).  
	After a large fraction of the disk gas loses its initial momentum during the first near head-on collision and stalls at the first contact surface,  pockets of dense, shock-compressed gas emerge ($t=50\; {\rm Myr}$ panels in Figure \ref{fig:1}).
	A few of these very dense clumps ($\Sigma_{\rm gas} \gtrsim 10^{3-4}\,{\rm M}_{\odot}\,{\rm pc}^{-2}$) rapidly cools and collapses to spawn clusters of stars before or while the DMDG's main stellar body forms \citep[$M_{\star,\,\rm SC} \gtrsim10^6\msun$ each; $t=70\; {\rm Myr}$ panels in Figure \ref{fig:3};][]{Kim2018, Madau2020}.   
	As the clumps continue to turn gas into stars in the next $\sim 50\;{\rm Myr}$, they are fed by the accreting cold gas while the frequent supernova explosions from young SFMCs eject the hot gas out of the new SCs ($t=110\; {\rm Myr}$ panels in Figure \ref{fig:3}; see also the bottom panel in Figure \ref{fig:4}). 
	The SCs consume or lose most of their gas by $t=160\;{\rm Myr}$, and continue to orbit around the DMDG until the end of our simulation at $t=800\;{\rm Myr}$.
	
	\begin{figure}[t]
		\centering
		\vspace{-1mm}     
		\includegraphics[width=0.485\textwidth]{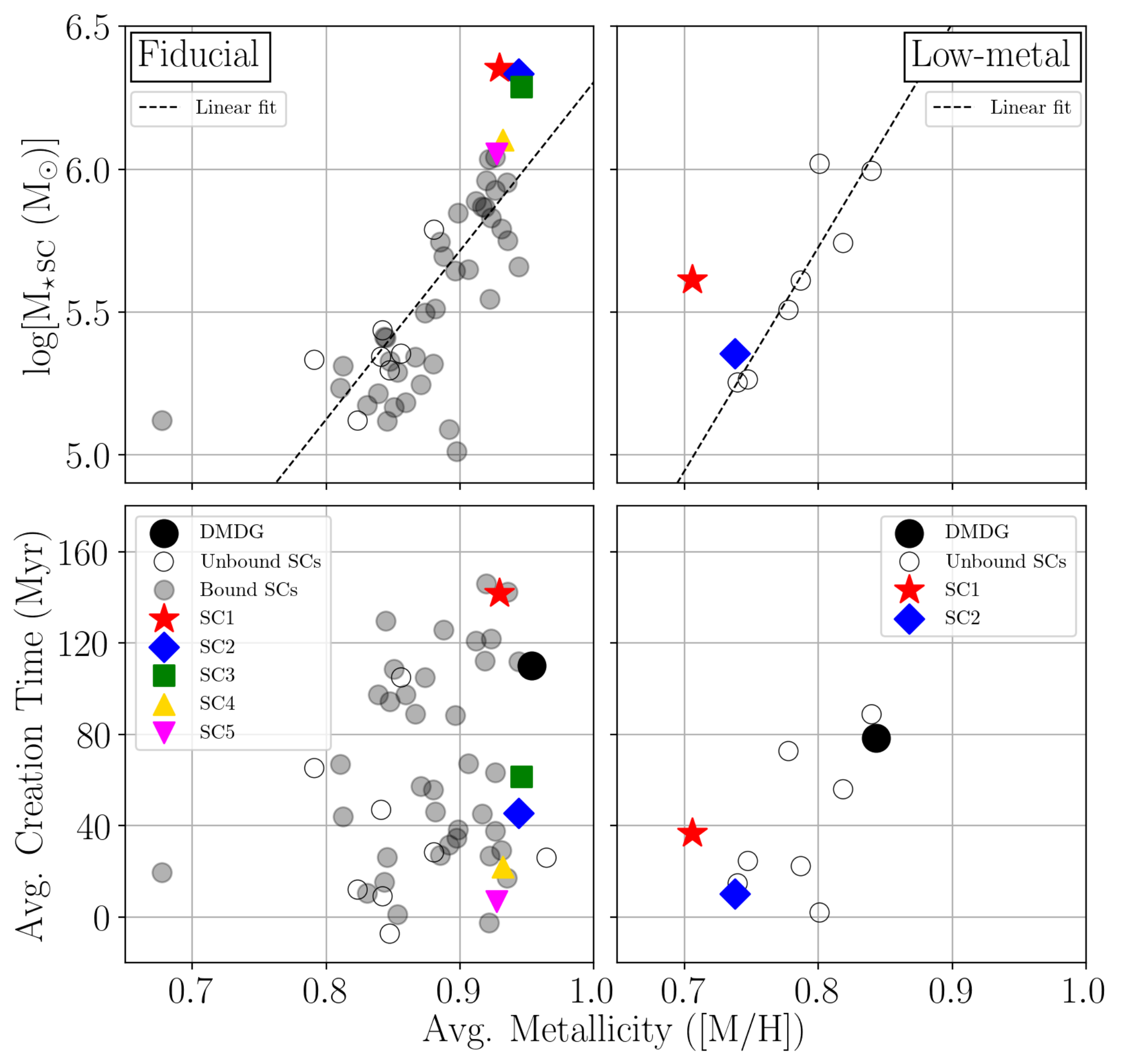}
		\vspace{-5mm}      
		\caption{
			Top row: the stellar masses versus the average metallicities of the SCs formed in the DMDG. 
			Bottom row: the average creation times versus the average metallicities of the SCs.
			The left panels are for the {\tt Fiducial} run (42 bound $+$ eight unbound SCs) and the right panels are for the {\tt Low-metal} run (two bound $+$ seven unbound SCs).
			Selected massive SCs are shown with the colored markers (the same SCs and markers as in Figures \ref{fig:2} and \ref{fig:4} for the {\tt Fiducial} run).
			The average metallicities of SCs are correlated with their masses.    
			We also denote the average creation time and metallicity of the DMDG as a whole with a black circle in the bottom panels.  
			In the {\tt Low-metal} run, the stars in the DMDG is more metal-enriched than the two bound SCs.  
			See Section \ref{sec:3.2} for more information. 
		}
		\label{fig:5}
		\vspace{0mm}      
	\end{figure}
	
	Figure \ref{fig:4} presents the star formation history in the most massive DMDG of the {\tt Fiducial} run and its five most massive member SCs.
	Most of the stars in the DMDG and the five SCs are born in $<150\;{\rm Myr}$ after the progenitor galaxies' pericentric approach ($t=0$–150 Myr), which is consistent with our earlier studies \citep{Shin2020}.
	After the initial burst of star formation, the DMDG maintains to form stars at a modest pace ($\sim 10^{-1}-10^{-2}\,{\rm M}_{\odot}\,{\rm yr}^{-1}$), while none of the massive SCs  form additional stars at a meaningful rate after $t=150$ Myr as they are nearly devoid of gas.
	The widths of star formation peaks for these SCs are $\sim$50  Myr, which are significantly narrower than that of the DMDG.  
	
	\begin{figure}[t]
		\centering
		\vspace{-1mm}      
		\includegraphics[width=0.44\textwidth]{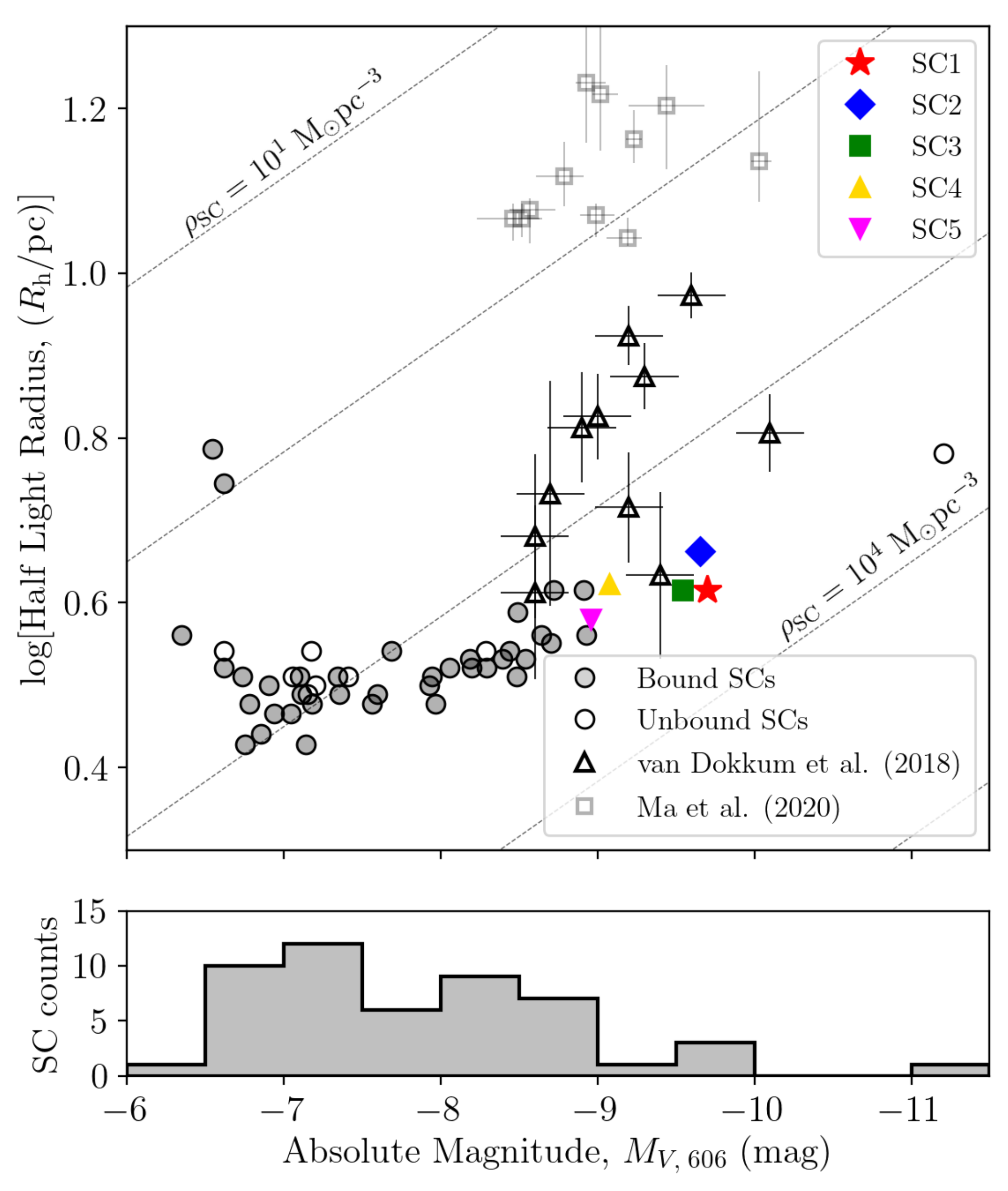}
		\vspace{-2mm}      
		\caption{
		    Top panel: the absolute magnitudes ($M_{V}$) versus the half-light radii of the SCs formed in the most massive DMDG in the {\tt Fiducial} run. 
			$M_{V}$ is estimated from the SC's mass, assuming the mass-to-light ratio $M/L_V = 2$.
			The five most massive SCs are shown with the colored markers (the same SCs and markers as in Figures \ref{fig:2}, \ref{fig:4} and \ref{fig:5}).
			We overplot the luminous GCs located in DF2 ($M_{V} < -8.5$; triangles from \citealt{vanDokkum2018b}; squares from \citealt{Ma2020}); here the half-light radii and $M_{V}$ in the F606W band are used. 
			See Section \ref{sec:3.2} for more information.
		}
		\label{fig:6}
		\vspace{2mm}      
	\end{figure}
	
	Then in Figure \ref{fig:5}, we make scatter plots of all the SCs  formed with the DMDG on the plane of the stellar mass versus the average metallicity, and on the plane of the average creation time versus the average metallicity.   
	The left panels are for the {\tt Fiducial} run and the right panels are for the {\tt Low-metal} run (note that the {\tt Low-metal} run with the initial metallicity of $10^{-2} \,Z_\odot$ produces nine SCs compared to 50 in the {\tt Fiducial} run; see Table \ref{tab:1} and Section \ref{sec:3.1} for more information).
	In the bottom panels, we also denote the average creation time and metallicity of the stellar body of the DMDG with a black circle. 
	Two observations could be made.  
	First, we find that the average metallicity of a SC correlates positively with its mass (top row) rather than with its creation time (bottom row). 
	We argue that because a massive SC creates more stars that consequently injects more metals into the ISM, and retains a higher fraction of supernova ejecta in its deeper gravitational potential well, the next generations of stars in the SC inherit higher metallicities from the metal-enriched ISM \citep[][]{Bailin2009,Fensch2014}.
	Therefore, the more massive a SC is, the higher its average stellar metallicity is. 
	Second, readers may note that in the {\tt Low-metal} run, the average metallicity of the DMDG is higher than that of the two bound SCs  (colored markers in the bottom-right panel).  
	A similar discrepancy is observed in DF2:  ${\rm [M/H]}\sim -1.07$ for DF2's stellar body and $\sim -1.63$ for its SCs  \citep{Fensch2019}.
	The higher metallicity in the DMDG's main stellar body could be attributed to the continuing star formation after the initial starburst.  
	The later generations of stars formed in the more metal-enriched ISM raise the average stellar metallicity of the DMDG, while the SCs' metallicities rarely change once the initial starburst ceases.\footnote{
    	Note that this trend is found only in the {\tt Low-metal} run, but not in the {\tt Fiducial} run where the initial gas metallicity is already set to $1 \,Z_\odot$. }
	
	Finally, we analyze the individual characteristics of the SCs, such as their luminosities.
	Figure \ref{fig:6} shows the scatter plot on the plane of the absolute magnitude ($M_{V}$) versus the half-light radius ($=$ half-mass radius $R_{\rm h}$) for all the SCs in the most massive DMDG of the {\tt Fiducial} run.
	The absolute magnitude of a SC is estimated from its mass, assuming a typical constant mass-to-light ratio $M/L_V=2$ \citep[e.g.,][]{vanDokkum2018b}. 
	The SC masses plotted in Figure \ref{fig:6} range from $1.2\times 10^{5}$ to $9.0\times 10^{6}\msun$.
	Among them, 32 gravitationally bound SCs are with $-8.5<M_{V}<-6.5$, while 10 bound SCs are with $M_{V} < -8.5$. 
	For comparison, we include the luminous SCs observed in DF2 \citep[$M_{V} < -8.5$;][]{vanDokkum2018b, Ma2020}.
	The absolute magnitudes of the luminous SCs are similar in our {\tt Fiducial} run and DF2. 
	Note that the peak of our cluster luminosity function (the bottom panel of Figure \ref{fig:6}) lies between the peaks of the Milky Way's GCLF and DF2's \citep{vanDokkum2018b}.\footnote{
    	84 (Milky Way) $/$ 4 (DF2) GCs are with $-8.5<M_{V}<-6.5$, and 20 (Milky Way) $/$ 11 (DF2) GCs are with $M_{V} < -8.5$.}
	We argue that the extremely dense, shock-compressed gas clumps induced by the high-velocity galaxy collision could give rise to multiple luminous SCs—a scenario considered by \citet{Trujillo2020, Trujillo2021} and demonstrated with simulations by \citet{Kim2018} and  \citet{Madau2020}. 
	Meanwhile, the half-light radii $R_{\rm h}$ of the simulated SCs are 1.5–3.0 times smaller than the estimated radii of the GCs in DF2.
	As was for $R_{\rm h}$ of our simulated DMDG, it is important to note that our simulated SCs evolved only for $< 800 \;{\rm Myr}$.
	If evolved for a sufficiently long time, the radii of SCs could change due to, e.g., the dynamical heating by their host galaxy.\footnote{
    	For example, \cite{Gnedin1999} showed that during orbital passages near the host galaxy, GCs could be dynamically heated by the host. We also note that the detailed dynamics of supernova blast waves or remnants is not fully resolved with 1.25 pc resolution (or SFMC particle of $\gtrsim 200 \,{\rm M}_\odot$), making it difficult to accurately estimate the disruption time scale and the inner structure of each SC. This warrants an even higher resolution simulation.}

	\section{Conclusion}
	
	The diffuse dwarf galaxies DF2 and DF4, with their alleged absence of dark matter and multiple luminous member GCs, have been a topic of intense debate as the astrophysical community tries to comprehend their formation mechanism.
	Using a suite of high-velocity galaxy collision simulations with 1.25 pc resolution, we have demonstrated that such a collision induces the formation of the DMDGs with $M_{\star}\sim10^8\msun$.
	With numerical resolution significantly improved from our previous study \citep{Shin2020}, we resolve the dynamical structure of the produced DMDGs and find that a co-planar prograde–prograde galaxy collision can produce a compact, rotation-supported, oblate DMDG.
	We have also resolved and studied the SCs forming simultaneously with the DMDGs, which later could become progenitors of GCs.
	A galaxy collision with a high relative velocity of $\sim 300\;{\rm km\;s}^{-1}$ creates pockets of severely shock-compressed gas that rapidly turn into multiple luminous SCs  ($M_\star \gtrsim10^6\msun$, $M_V\lesssim-8.5$ mag) that are in line with the observed GCs in DF2.
	The majority of stars in the DMDGs and their SCs  form less than $150\;{\rm Myr}$ after the progenitor galaxies’ pericentric approach.  
	Many properties of the observed DMDGs' GCs are reproduced.  
	In our {\tt Fiducial} run, 10 luminous, gravitationally bound, virialized SCs  have a line-of-sight velocity dispersion  $11.2\;{\rm km\;s}^{-1}$.   
	In the {\tt Low-metal} run with the initial gas metallicity of $10^{-2} \,Z_\odot$, the DMDG shows a higher average stellar metallicity than the SCs.  
	
	Our study suggests that a high-velocity galaxy collision is a promising candidate for the simultaneous formation of DMDGs and their luminous member GCs.
	In the forthcoming studies, we aim to examine the galaxy collision-induced DMDG formation scenario with more sophisticated subgrid physics that fully takes into account the evolution of interstellar medium due to energetic stellar feedback with even higher resolution (see footnote \ref{footnote:feedback}), in a cosmological context for $\sim10$ Gyr, comparable to DF2's age \citep[$\sim 9$ Gyr;][]{vanDokkum2018b, Fensch2019}  designed to follow its later evolution.
	
	\vspace{2mm}
	
	The authors would like to thank Jérémy Fensch, Yongseok Jo, Taysun Kimm, Myung Gyoon Lee, Boon Kiat Oh, Joseph Silk, Sebastian Trujillo-Gomez, and Sukyoung Yi for insightful discussions. 
	Ji-hoon Kim acknowledges support by Samsung Science and Technology Foundation under Project Number SSTF-BA1802-04, and by the POSCO Science Fellowship of POSCO TJ Park Foundation.
	His work was also supported by the National Institute of Supercomputing and Network/Korea Institute of Science and Technology Information with supercomputing resources including technical support, grants KSC-2019-CRE-0163 and KSC-2020-CRE-0219.
	The publicly available {\sc Enzo} and {\tt yt} codes used in this work are the products of collaborative efforts by many independent scientists from numerous institutions around the world.  
	Their commitment to open science has helped make this work possible.
	
	\software{
		{\tt yt} \citep{Turk2011},
		{\sc Enzo} \citep{Bryan2014,Brummel-Smith2019},
		the {\sc Grackle} chemistry and cooling library \citep{Smith2017},
		{\tt numpy} \citep{oliphant2006guide},
		{\tt scipy} \citep{Virtanen2020},
		{\tt matplotlib} \citep{Hunter2007}
	}

	\bibliographystyle{aasjournal}
	
	
\end{document}